# Would You Mind Being Watched by Machines?
# Privacy Concerns in Data Mining

Vincent C. Müller
Princeton University, Hellenic Studies
Anatolia College/ACT
www.typos.de

**Abstract:** "Data mining is not an invasion of privacy because access to data is only by machines, not by people" – this is the argument that is investigated here. The current importance of this problem is developed in a case-study of data-mining in the USA for counterterrorism and other surveillance purposes. After a clarification of the relevant nature of privacy, it is argued that access by machines cannot warrant the access to further information, since the analysis will have to be made either by humans or by machines that understand. It concludes that current data-mining violates the right to privacy and should be subject to the standard legal constraints for access to private information by people.

> *"We have to find the right balance between protecting our security and protecting our liberty. If we fail in this effort by drawing the line in the wrong place, that is, overly favoring liberty or security, then the terrorists win and liberty looses in either case."*
> Michael Hayden, Director NSA 1999-2005,
> Director CIA since May 2006
> (Hayden 2002, 11)

## 1. Introduction: The Problem

In December 2005, US President George Bush admitted that he had ordered the National Security Agency (NSA), in the wake of the Sept. 11th attacks, to "listen in" on communications between the USA and abroad, and also within the USA – without obtaining relevant court orders. The NSA had previously been responsible only for spying on foreign soil or on foreign spies within the US.[1] This admission sparked a fairly intense debate about the justification of the move, and several prominent supporters of Bush's policy appeared in the media. Former Attorney General Edwin Meese said, amongst other things: "We are not talking about wiretapping. … This isn't eavesdropping, it is just surveillance." He further alluded that "certain technical procedures" were used, but that these are "classified". Another former member of republican administrations, Charles Fried, asks what harm is done by such activity: "Is a person's privacy truly violated if his international communications are subject to this kind of impersonal, computerized screening?" Judge Posner is explicit "But machine collection and processing of data cannot, as such, invade privacy." (2005).[2] What is argued here falls into a pattern that is increasingly popular recently: we just have machines, computers, analyzing data, nobody is listening in, nobody intrudes into your privacy, so don't worry, there is no problem. – Is this correct?

---

[1] "The law requires the NSA to not deliberately collect data on US citizens or on persons in the United States without a warrant based on foreign intelligence requirements." (9/11 Commission 2004, 87) This avoidance of domestic data was considered a significant factor for the failure to prevent the Sept. 11th 2001 attacks (ibid.).

[2] Edwin Meese on MSNBC, in Chris Mattews' program "Hardball", Jan 12th, 2006, 19:45. Fried in his article in the *Boston Globe* (Fried 2005). Fried was solicitor general in the second Reagan administration. Taipale argues data-mining is "…different than claiming that 'everybody is being investigated' through pattern-matching. In reality only the electronic footprints of transactions and activities are being scrutinized." (Taipale 2003, 66).

This line of argument is used to defend practices that may constitute a very significant and yet somewhat underrated threat to privacy: Blanket surveillance of whole populations. Unlike traditional wiretapping etc. of particular individuals, this activity is under very limited, if any, legal control – while offering very powerful possibilities. It is dependent on computing technology since both the source data and the tools of analysis must be digital. The potential result is a very detailed picture of the activities of persons and groups, that can serve all sorts of purposes, from crime prevention and detection to political control and targeted advertisement.

A practical question is how much of this is happening and who is doing it. The theoretically interesting question is whether we should mind. Much of this data may not be considered private and is thus not legally protected (while my mobile phone conversations are covered, the location of my phone or my walking down a particular street might not be). Also, to be considered a violation of privacy, it is important *who* or rather *what* it is that has access. We do not mind if what has access is just a machine, rather than a person: It would seem silly to worry about a computer at the phone company that processes call minutes and prints a bill at the end of the month. So, if a computer of the US National Security Agency (NSA) processes the same calls, why worry? If you worry, is that because you have something to hide? Would there be reason to worry even if the intentions of these agencies were entirely benevolent?

### 1.1. The Argument in Brief

I propose to investigate this question via an argument that runs as follows: (1) The data of US citizens (and others) is systematically analyzed by data-mining techniques. (2) An analyzing machine either understands what it analyzes or it does not. (3) If it does understand, privacy has been breached. (4) If it does not understand, the analysis must be completed by a human being, so privacy is breached. (5) Our privacy is breached by data-mining. Since this argument has a valid logical form (it is what is called a "constructive dilemma"), the only question is whether the three premises of this argument are true.

The argument above does not deal with the question which information should be private, to whom, and under what circumstances. It just clarifies whether the use of "data-mining" alters the privacy situation, in particular whether the use of machines could fail to violate privacy while conventional "reading" by a human would. It also argues (in premise 1) that this is not a theoretical problem, but ac-tually occurs. Finally, it argues (6) that the breach of privacy in data-mining cannot be warranted by the results of data-mining themselves: If data-mining tags an individual as suspect, that tagging is not sufficient legal grounds for breaching the person's privacy.

## 2. Data-Mining and Artificial Intelligence

### 2.1. What Could Be Done

In the old days, before 1989, the German Democratic Republic (GDR) secret service "Stasi" steamed open letters and employed people to read these. Looking through all the mail from the West thus took an enormous amount of time and manpower – so while all incoming mail was checked, outgoing mail was not checked completely. If this information had been available digitally, it would have been cheap and fast to store it and to process it all. This is the situation now. The content of our *communications* is digital: e-mail, sms, mms, www, telephone in its various forms, videophone, … The same applies to almost all *personal information* systematically gathered for other reasons: tax records, state registries of various sorts, health records, insurance records, bank and credit card records, employment records, police and other legal records, customer cards records, organization records, phone connection data, library loan records, purchase records, cable TV logs, web site logs, web site "cookies", web site contents, mailing list logs, search engine databases, … In addition to this, there may be personal information that is collected specifically for some purpose of *surveillance*: mobile phone locations, GPS in vehicles[3] and on humans, transponders in vehicles (for road tolls etc.), secret coding of color-photocopies and other output devices,

---

[3] This is a common form of vehicle theft prevention worldwide. Germany now requires all trucks to carry GPS and logs their travel to collect motorway tolls. In 2005, it was seriously considered in Great Britain to control all movement of all vehicles via GPS, as a means to collect road tolls for the entire country, in replacement of road tax. Access to the center of London is already controlled by a system that logs license plates of all vehicles entering and leaving the designated zone. GPS has been used by US car rental companies to issue speeding tickets to their customers, a practice that has been legally challenged (O'Harrow 2005, 292).

RFIDs[4] and other forms of electronic tagging, cameras in public spaces[5] and on satellites, x-ray cameras (used in customs), spyware on PCs and in networks, ICT Implants, DNA detectors, digital dust sensors, software for recognition of individuals from text, audio or video data …[6]

All of this data can now be stored forever. The resulting massive databases can be combined and analyzed with intelligent data mining and other computing techniques. Note, however, that the amount of data involved here is staggering, easily several magnitudes higher than all what is stored on the WWW – after all, most of the data that is computer-stored world-wide is personal data for some person. It is estimated that each human generated an average of 250 MB of digital data each year in 1999, but 800 MB in 2002, i.e. about 5 million terabytes each year overall – this number will continue to rise sharply with the spread of information technology.[7]

The NSA director warned in 2002 that all this data would be very hard to analyze: "This year, the world's population will spend over 180 billion minutes on the phone *in international calls alone*." (Hayden 2002, 6) He also stresses the enormous increase in data to be processed, due to new technologies, such as the mobile phone, Internet use, etc. Note also that this data will be in many languages, in some cases in more than one language in one communication. Note also that the basic analysis must be done in "real time", i.e. as fast as the new data is coming in.

---

[4] Now spreading very widely: Used by all suppliers of WalMart, i.e. practically all consumer products in the USA. Required by law in the EU to be injected in all domestic cats and dogs.

[5] It is estimated, as an extreme case, that London has 2.5 million video cameras, the average Londoner is filmed 300 times per day (US Congress 2002, 2). The whole USA had about 2 million cameras in 2002 (Bailey 2004, 75; Keenan 2005, 57). Some London boroughs connect their video data to face-recognition software (O'Harrow 2005, 165f).

[6] For some recent research by DARPA and others on the identification of people and vehicles in video and audio data, see (IEEE 2003). For face recognition, see (O'Harrow 2005, ch. 6). (All applicants for US-visa and all "immigrants" entering the USA are photographed and fingerprinted since 2004.)

[7] The sequence by 1024 is megabyte, gigabyte, terabyte, petabyte, exabyte, zettabyte, … The original report explains "If digitized with full formatting, the seventeen million books in the Library of Congress contain about 136 terabytes of information." (Lyman and Varian 2003) Such numbers must be taken with great caution, particularly since the notion of "amount of data" is utterly meaningless if data is not taken to be digital. (How much data is there on your desk?) Even digital data can be compressed and have many formats. Some sources say that "intelligence data sources grow at the rate of four petabytes a month" (O'Harrow 2005, 212).

### 2.2. How to Analyze: Types of Data Mining

In the old days, if someone wanted to listen to your telephone conversation, they had to get physical access to the wire connecting you to the other person, they had to "tap the wire", as the expression goes (even if the information was already digital). Today, a very large part of communication is wireless at some point: either because your communication device is not connected by wire (a cordless phone, a mobile phone, a wireless PC), or because the communication is transferred wireless somewhere on the way, often via satellite. This is where it can easily be intercepted. Also, of course, traditional wire-tapping has its role in the case of "wires" that connect networks, countries, etc.

The digital data that is intercepted and analyzed can have all sorts of contents, numbers, text, audio, images, video, etc. Most of this data is then converted to text. Text has the advantage that it is per se digital, normally in characters, which reduces file size and facilitates analysis significantly.

Now, if one had a nicely structured database where all data is organized in fields and properly tagged, one can write "queries" that will solve all problems – but of course the data gathered by interception does not come with much structure (though e-mail, for example, has standard header information). This requires the use of the techniques known as "data mining", i.e. algorithms designed to discover patterns in data sets, where the data sets were not already designed to yield these patterns. (In a database that has full names in one field and whole address in another, you cannot formulate a query to find all persons whose first name is "Mohammad" and who live in Philadelphia. In an e-mail text, you need to use clues to find out which are the names, and which of these is the name of a place.) Data-mining can thus be done on all sorts of data, not necessarily on structured databases, and typically not on uniformly structured data. So, for example, the connection of seemingly unconnected matters that a company like amazon.com does for its customers does *not* rely on data-mining techniques, but rather on a well-organized database. The quality of output depends critically on the quality and organization of input, so the correcting and standardization of data is an important first step. (Otherwise, the old computer saying "garbage in, garbage out" applies.)

There are basically three types of analysis relevant here, "subject-based", "matching" and "pattern-based".

*Subject-based* analysis is a form of targeted personal tracking (Bailey 2004, 103), a continuation to traditional police work that starts with a suspect and trying to find out more about the person, trying to link him/her to other people who might be involved, for example. It presupposes that a particular person has been identified as a suspect – so, in principle, it should be subject to the standard regulations concerning the gathering of information about that person, of personal information.

*Matching* concerns personally identifiable information that is matched with other identifiable information (e.g. in a different database) to catch suspicious behavior, such as receiving loans on one's home from several different banks, or receiving social security and a salary at the same time. This is a technique which is most interesting for commercial application. Matching requires integration or comparison of data from different sources, so it raises the concern that data will be used for a purpose the person concerned has not agreed to, or does not even know about. It is essentially just a use of the standard techniques for the discovery of inconsistencies in databases, so it is normally not data mining. It requires that the data concerned can be allocated to an identifiable person, so, again, it is subject to the standard regulations.

*Pattern-based* analysis, finally, is the search for any kind of pattern in the data that might produce some information that is considered "interesting". Since it is often entirely unclear at the point in time when information is collected, which information will be considered interesting later on, this kind of analysis is the main domain of data-mining. One classical example may be the attempt to find out who had significant benefits on the stock markets from the 9/11 attacks; benefits that would indicate that this person knew that the attacks would take place. The available data was not collected and structured for this purpose, but now *this* was the interesting information. Another, much more common one, is the discovery of credit card fraud: If a transaction is done with your credit card that is "unusual", then the bank might contact the customer to make sure no fraud is involved. In police work, pattern-based analysis can be the continuation of the crime-detection technique of "profiling" suspects, popular since the 1970ies. (E.g., in California, private home owners below a certain age with an unusually high electricity bill were targeted by police under the suspicion that these home owners might grow marihuana in their basement. Young people who rent large trucks in the US and use c/o addresses have been investigated as suspects for terrorist attacks (DeRosa 2004, 8)).

A special case of pattern analysis is "link analysis" where a person is checked for their connections to other persons, e.g. their telephone conversations, bank accounts, locations, etc. (It turned out that all 9/11 hijackers had such links.)

Pattern analysis is probabilistic, that is it will try to allocate a probability for belonging to the correct output set, given a number of weighed factors. If the overall probability crosses a defined threshold, the data is considered a positive. The identifications suffer from a margin of error and the question typically is what is acceptable for a particular task: that the program identifies something as a match which is not (a false positive), or that the program fails to identify something as a match which is (a false negative). For the analysis of credit card data, a false positive is acceptable, but a false negative is a problem. On the contrary, for the filtering of unwanted "spam" e-mail, a false negative (unwanted mail that gets through) is acceptable, but a false positive (wanted mail that is blocked) is unacceptable. Programs that search for an extremely rare but highly dangerous activity (e.g. terrorists or politically undesired persons) must be constructed in such a way that they primarily avoid many false negatives, i.e. they will get false positives: innocent people will be caught in the wide net. In a second step, the false positives need to be narrowed down: It is no use being told that more than 200,000 people are suspected to belong to terrorist groups since this means that nearly all of these suspects are actually innocent.

The filters for such patterns can be constructed "top-down" if one has a specific pattern that is considered suspicious. It can also be constructed "bottom-up", by giving the program a "training set" of correct output results and asking it to find the pattern that separates this set from the others (DeRosa 2004, 11). The bottom-up procedure is potentially able to produce more unexpected results, but it is very hard to find out how relevant the pattern identified really is to the desired output. (Perhaps all suspects in the training set shared a feature that is irrelevant to their being suspects.)

Given the amount of data involved, it is safe to say that pattern-based data-mining will be used on personally identifiable information, such as credit card records, driver's licenses, tax filings, phone-call records etc. It is less efficient for surveillance, e.g. the analysis of the *contents* of communication or the recordings of "security cameras".

**2.3. What Is Known to Be Done**

Before we enter into some illustrative details, let it be clear that what is known to be known is very basic, and what is thought to be known is mostly speculation. The relevant activities are carried out by police, secret services and businesses. None of these want the public to know about their activities. Even much of the academic research in the area is classified. So, what we know are the very general official pronouncements and the occasional unintended news that surfaces.

It should be stressed, however, that in the US there are many companies that deal with data which in the European Union would be considered private. This applies to all imaginable sorts of customer data, but also to state-collected records and the cross-referencing of these (see Markle Foundation 2003). Specialized companies identify people's credit standing, name, address, phone numbers, height, weight and social security number (DeRosa 2004, 10). Much of this potential has been used by police and secret services after the 9/11 attacks, esp. in the "Matrix" program (see O'Harrow 2005, chs. 2 and 4). All of it is used by private companies for all sorts of purposes.

Some of the major companies involved at present are Axicon, ChoicePoint and LexisNexis – the latter now owns Seisint and is part of UK-based Reed-Elsevier. ChoicePoint alone has about 17 billion public records, 250 TB of data. It acquires data world-wide. In 2001, ChoicePoint offered information to the US government about all South-American people, e.g. Mexican voters with name, address, ID no. and date of birth (O'Harrow 2005, 145, 152). The offer was accepted. These companies work for all major corporations in the US to support marketing, customer-service, product development, employee screening, risk assessment, access control, fraud-detection etc. Some major software companies, such as Google, have their own data-mining techniques for internet searchers, webmail users, etc. (some indications on http://www.google-watch.org/).

2.3.1. Privacy for Us, Surveillance for Them

Very little of this activity is the result of directed, legally controlled surveillance of particular people. Most of it is carried out by secret services on foreign nationals. If a state A spies on the nationals of state B, the laws of state A typically do not disallow this, and the danger of state B complaining is small, given that it has no evidence and is possibly doing the same to the nationals of state A. (In the case of "Echelon", see below, it may well be that states A and B exchange information about the nationals of each, thus avoiding responsibility for spying on their own nationals (Keenan 2005, 43).)

In the following, I shall focus on the case of the USA. The official 9/11 commission acknowledges the difference between what is permitted domestically and what is permitted otherwise: "The FBI's job in the streets of the United States would thus be a domestic equivalent, operating under the U.S. Constitution and quite different laws and rules, to the jobs of the CIA's operative officers abroad.", continuing "The FBI is accustomed to carrying out sensitive intelligence collection operations in compliance with the law." (9/11 Commission 2004, 423). The Dept. of Defense TAPAC report proposes constraints based on privacy, but adds: "We recommend *excluding* from these requirements data mining that is limited to foreign intelligence that does not involve U.S. persons." (their italics TAPAC 2004, x).[8]

According to official data, the US operates some 15 different secret agencies (9/11 Commission 2004, 410). Prominent for our purposes are the CIA, which "collects, analyzes and disseminates intelligence from all sources", the NSA, which "intercepts and analyzes foreign communications and breaks code"[9], the Geospatial-Intelligence Agency (GSA), which "provides and analyzes imagery and produces a wide array of products, including maps, navigation tools, and surveillance intelligence", finally the National Reconnaissance Office (NRO), which "procures, launches and maintains in orbit information gathering satellites that serve other government agencies." (9/11 Commission 2004, 86f).

Given the secrecy and power of such agencies, one might get the impression that they know everything and can do anything – but we should be careful not to give in to this temptation. There are many well documented failures where US agencies did now know about important events before they occurred: e.g. the fall of the Berlin Wall 1989, nuclear weapon tests by India and Pakistan 1998, attacks on New York and Langley 2001[10]; they also failed to identify the whereabouts of

---

[8]  The same provisions, excluding military operations and intelligence activities overseas or against non-US citizens are in the "Department of Homeland Securities Appropriations Act", sect. 8131 (b), as quoted in (Taipale 2003, 10, n. 28).

[9]  For more information about the NSA, consider (Anonymous 2005).

[10]  About the failure to identify Sept. 11th: "To put this into perspective, throughout the summer of 2001 we had more than 30 warnings that *something* was imminent. We dutifully reported these, yet none of these subsequently correlated with terrorist attacks. The concept of 'imminent' to our adversaries is relative; it can mean soon or simply sometime

Saddam Hussein in 2002, or of Osama Bin Laden and other Al Queida operatives in recent years.

### 2.3.2. Surveillance and Terrorism

With the waning of the cold war and particularly after the attacks on Sept. 11$^{th}$ 2001, the perception in the USA as to who are the potential enemies that should be observed has changed significantly. The target is now "terrorism", which is far less specific than particular foreign states or foreign military – but still couched in the same terms as a "war". It involves groups or even individuals who may well not have identified themselves as enemies before they attack. Accordingly, the techniques for identifying such enemies, preventing their attacks and persecuting perpetrators would have to adjust (e.g. Markle Foundation 2002). One cannot expect few, rich sources of information, as may be the case in traditional espionage, but has to make sense of many small pieces of information. Having said that, the NSA publicly admitted that before 9/11 (!) "FBI headquarters routinely receive about 200 reports daily from us." But adds "… we have been more agile in sharing information with some customers (like the Department of Defense) than we have with others (like the Department of Justice)." (Hayden 2002, 9). There was supposed to be information sharing from the FBI to the intelligence services, but not in the opposite direction. This obstacle was frequently blamed for the failure to identify 9/11 (sometimes called "the wall") and US Congress suggested "… to develop capability to facilitate the timely and complete sharing of relevant intelligence information both within the Intelligence Community and with other appropriate federal, state, and local authorities." (quoted in Taipale 2003, 5).

Allow me to mention some known programs:

In 2002, the US government proposed TIA, the *"Total Information Awareness"* system, in which all US bank records, tax filings, driver license information, credit card information, financial records, medical data, telephone and e-mail records, travel information and other data would be combined under the auspice of DARPA – who had been working on the concept since the late 1990ies (Rosenberg 2004, 390f; O'Harrow 2005, ch. 7). These databases were to be augmented with conventional intelligence, improved human identification, speech-to-text (EARS: effective affordable reusable speech to text), automatic translation technologies

---

in the future." (Hayden 2002, 4) Hayden also stresses the difficulty of identifying and processing several languages and the crucial factor of processing on time.

(TIDES) and analyzed with data-mining technologies (in GENISYS). In one 2002 presentation, the "automation goal" was described as "read everything without reading everything" (Armour 2002). TIA is deeply embedded into various DARPA computing technology projects (DARPA 2003, 5-11; IAO 2003). Following widespread criticism, the program was renamed "Terrorism Information Awareness" in March 2003 but had to be dropped due to the continuing criticism and a forthcoming damning report of the Department of Defense (TAPAC 2004).[11] The order to cancel such activities was signed by President Bush in October 2003, but contained a "classified annex", specifying for which purposes the continuation of the program was permitted (Taipale 2003, 10 - cf. 39ff). – What was shocking to the US public was the use of private US data. What is clearly continued is the use of private data from abroad and from non-US citizens. This clearly includes data gathered by security services in violation of local law.

A similar program was *"Matrix"* (Multi-state Anti-Terrorism Information Exchange), which aimed to combine these publicly available identifiable records and, when joined by the FBI, added criminal records, driver license photos – and all that through a private company with experience in commercial data-mining: Hank Asher's *Seisint* (O'Harrow 2005, 107f, 121ff). Though police was impressed with the new abilities of this "one stop-shop", public concern about this combination of criminal and commercial records forced the cancellation of US-wide Matrix late in 2003 when more and more US states refused to cooperate. However, some states still appear to work with the system (Taipale 2003, 16 still hopes for its nationwide extension). O'Harrow says: "Make no mistake, though, Asher's technology will be there, working behind the scenes, no matter what it is called, and now quite possibly on a global scale." (O'Harrow 2005, 124). Bailey deplores the cancellation of TIA, but also offers this consolation: "Going forward, it's likely that many of the ideas and much of the technology behind TIA will live on and be utilized by different agencies, …" (Bailey 2004, 106). Given the "Classified Annex" mentioned above, this is not just speculation, but certainty. Also, the founding of the National Counterterrorism Center may point in this direction.

The 2001 "Patriot Act" required many companies and organizations in the USA to report "suspicious activity". The *Financial Crimes Enforcement Network*, FinCEN, is now used by banks to report "suspicious" customer behavior to the FBI – about 300.000 reports were filed in 2003 alone. FinCEN also shares its entire database with the FBI each month. Customers are not informed (O'Harrow 2005, 266).

---

[11] Most references to TIA have been removed from the DARPA sites, but Director Pointexter's outlook can be gathered from his slides (Poindexter 2002).

*Echelon* is a network of secret services operated by Australia, Canada, New Zealand, the United Kingdom and the USA, which is thought allow near-total surveillance of communication by potentially intercepting all wireless communications worldwide. The Soviet Union had a similar network, called "Dozor", which is currently being updated (Keenan 2005, 42f).

Everyone who boards a flight to or inside the US has been checked by *"Computer Assisted Passenger Pre-Screening"* (CAPPS), installed in the late 1990ies and updated in 2003 to CAPPS II, run by the Transportation Safety Agency (TSA), a division of the Department of Homeland Security. Any passenger's name, address, US address, date of birth and US phone number is required. TSA will query various databases and classify passengers as "green", "yellow" (special scrutiny) or "red" (inform police). CAPPS does not mine data, it matches entries from many databases it queries (Taipale 2003, 37ff).

The FBI has a system called *"Carnivore"* which it installs in the computers of Internet service providers. The system reads the headers of all e-mails sent and received, as well as the IP addresses of all communications, such as web sites, ftp sites, etc. If a communication matches the search criteria, its content is stored and sent to the FBI. It is supposed to catch only information relevant to a particular suspect, and to be subject to the authorization by a court order, including the secret FISA courts (Keenan 2005, 71; O'Harrow 2005, 257). Similar systems are in use in other countries as well.

### 2.4. Summary

Given this quick survey of known surveillance activities by US services and businesses, it seems that we have established our first premise, that data-mining on our data is actually taking place – on communications, databases and surveillance data. I presume that it is granted that the machines either understand or do not understand what they analyze, even if there may be a grey zone here, so what we need to see now is whether this constitutes a violation of privacy.

### 3. The Nature of Privacy and Its relation to Understanding

In order to establish that understanding the data would imply that privacy is breached (3), we need to gain a basic understanding of what constitutes "privacy" and how it relates to "understanding". It is surprisingly difficult to find a standard definition of privacy in the literature, or even a standard set of definitions.

One thing we do know is that it is very closely connected to the notion of a "person", particularly the natural person, but also the "legal person", i.e. the corporation or organization. I follow the classic treatment in (Rosenberg 2004, 349f) in distinguishing three kinds of privacy, that I would call 1) *personal space*, 2) *personal body* and 3) *personal information* – a person would have a right to privacy in these three domains. The one currently under threat by computing technology is that of personal information. "Personal" information must be information about that same person, but which of this is sufficiently personal to be considered "private" will be different with different individuals and societies – and perhaps even situations.

A classic formulation is, in this vein: "Privacy is the claim of individuals, groups or institutions to determine for themselves when, how and to what extent information about them is communicated to others." (Alan F. Westin 1967, 7, quoted from Rosenberg 2004, 349).

It is best to consider this as a right with no inherent bounds. In principle, a person can consider any personal information "private", and refuse to disclose it. For example, if I think "what my face looks like" is information I do not wish to disclose (I consider my face a "private part", or I am ashamed of its appearance), I have the right not to disclose it. Having said that, if this broad definition is accepted, living with other persons necessarily involves disclosure of private information. It is a matter of societal and personal negotiation, how much personal information one is willing to disclose, and to whom. Many social interactions will presuppose the disclosure of private information, often to the extent that one does not realize that this is the case and does not consider the information personal. For example, in order to get married or to open a bank account in the US, one has to disclose one's face and one's address. If one refuses to disclose this information, the other part will refuse to engage in the interaction. If one is a woman in Kuwait, however, one might be able to carry out both these interactions without disclosing one's face, but one would have to disclose one's address as well.

What we call "private information" then, is the personal information that we do not *wish* to disclose to some people. Whether or not we have that *right* depends on the arrangements that we make with those other people, or with our societies at large. Deviations from the social norm, wanting to hide very much or very little information, are not considered a sign of good psychological health – but it is still a right, and presumably few would want to go to the extremes and hide all or hide nothing.

It should have become clear that we are talking about a *right* of privacy, not about a *value*. This is frequently overlooked when one talks about balancing privacy with other desirable values, such as a safe society, efficient police-work, etc. – but privacy is not just one utility that is to be balanced against others. As a right, it cannot be violated other than by the individual consenting or else, by the society agreeing on rules under which circumstances this right can be violated (as it does with other rights). This will typically take the form of laws that specify who can violate the right and under which circumstances, e.g. who may listen in on telephone conversations, exercise violence, or imprison a person. Such laws will invariably state that the violation of rights is permitted only to agents of the state.

As a general guideline to what is considered reasonable in "Western" societies, it will be useful to quote the classic guidelines established by a 1973 report to the US Secretary of Health:

1) "There must be no personal-data record keeping system whose very existence is secret.
2) There must be a way for an individual to find out what information about him is in a record and how it is used.
3) There must be a way for an individual to prevent information about him that was obtained for one purpose from being used or made available for other purposes without his consent.
4) There must be a way for an individual to correct or amend a record of identifiable information about him.
5) Any organization creating, maintaining, using or disseminating records of identifiable personal data must assure the reliability of data for their intended use and must take precautions to prevent misuse of the data." (HEW 1973)

Given these clarifications, it is clear that privacy concerns the access of information by other *persons*. A necessary condition for a violation therefore is that the information is *understood* by someone.

### 3.1. Is Data-Mining per se Wrong?

3.1.1. New Patterns

It is definitional for data-mining that it searches for patterns that were not foreseen in the structuring of the data. Now, given the definition of the right to privacy above, any non-authorized use of personal data would constitute a breach of privacy that demands legal authorization. So, if the data is searched for *unforeseen* patterns, for patterns that the individual concerned cannot yet have agreed to be used, this would seem to show that *all* data-mining constitutes a breach of privacy. This does not follow, however. It may well be that the authorized use of the data covers unforeseen searches as well. For example, it may well be that a credit card company has been authorized to take all measures that prevent fraud, and it discovers that a certain pattern indicates fraud. Searching for that pattern would then be covered by the authorization. – The conclusion should thus not be that all data-mining violates privacy, but that all data-mining that is not implicitly or explicitly authorized by the persons concerned does.

### 3.1.2. Access to the Data vs. Use of the Data

If I give my phone company the permission to store connection data to write the monthly bill, I do not thereby allow them to perform any other analysis on the data. This is a new use, so the mere performing of the analysis is subject to my permission. If I deny that permission, the phone company can conclude that this denial makes the continuation of our relation undesirable. – Note that the problem starts with access to the data. Once someone has access and a use for some analysis, it will be hard to prevent them from privacy-violating new uses.

These kinds of violations have become a significant problem with the development of new data-mining technologies that allow the finding of patterns which had not been anticipated. For example, if one could find *all* photographs on the Internet that show your face and identify them as showing you, and perhaps with whom, and doing what. The same issue arises when data-mining can produce new information from data that is accessible to an organization, e.g. the phone connection records to my mobile telephone company, or the tax statements to the ministry of finance. This problem is particularly urgent in the cases where data that is originally anonymous can be allocated to particular individuals by technical means.

Some of this problem might be alleviated technically, e.g. by preventing identification, in the new field of what is called "Privacy Preserving Data Mining" (see Kargupta *et al.* 2004; Vaidya, Clifton, and Zhu 2005), in which some see the solution to all problems (Taipale 2003) – while what it can do, at best, is to show that not all data-mining involves a breach of privacy. This, however, was never in doubt. In this context, it is worrying if some suggest that there is a "liberating" use of data-mining techniques without a user's consent, as it happens in the "Carnivore PE" project for a "toy" version of the FBI software, that is let it loose on networks with no authorization. The "… goal was to extend the FBI software, improve it, and in so doing inject a new design philosophy into the technology. The hope is that it will both increase public outrage over the excesses of data surveillance and also increase public awareness of those same technologies." (Alexander 2006, 485).

### 4. Understanding in Machines

To cut a very long story short, it is almost universally agreed that current computer systems do not understand, in particular they do not have *intentional states*, states that are *about* something in the world (for an introduction, see Crane 2003, esp. ch. 3). So, when they find a particular pattern, they cannot know what that pattern means. John Searle has presented an influential argument that *no* computer will be able to understand, at least not in virtue of being a computer (Searle 1980; cf. Preston and Bishop 2002). This raises the question whether our fourth premise is correct:

"(4) If it does not understand, the analysis must be completed by a human being, so privacy is breached."

How far will an automated analysis without understanding go? Will statistics distinguish the guilty from the suspicious (the terrorist from the philosopher who analyzes problems of privacy)? The experts' verdict is quite clear. DeRosa summarizes her view: "These techniques … are not likely to be useful as the only source for a conclusion or decision." (DeRosa 2004, v). Taipale, generally a supporter of data-mining, says that "a guiding principle … should be that data mining not be used to automatically trigger law enforcement consequences" (Taipale 2003, 32). It lies in the nature of data-mining for security, that it will only produce a certain probability for a match, but also many false positives. Unless a human investigator is involved at this stage, who views the data, the analysis remains useless. But at this stage the fact that someone is considered "suspect" by the system does not mean he/she is sufficiently suspect to warrant a breach of privacy, e.g. a court order.

Reports in the press indicate that problem of false positives is currently acute: *The Washington Post* headlined "325,000 Names on Terrorism List: Rights Groups Say Database May Include Innocent People" (Pincus and Eggen 2006). Officials from the National Counterterrorism Center (NCTC) responded by pointing out that a number of people may be entered with several names and spellings, thus reduc-ing the number to more than 200,000 actual people. Some 32,000 people on this list have been coded as "armed and dangerous" but have been given "the lowest handling code", meaning FBI is not informed if they are encountered. In other words, suspicions are not substantial enough. The Washington Post picked up neither on this self-confessed indication of incompetence (1/3 of names are misleading) nor on the obvious irony of "may include innocent people" in the title. Ignoring that most almost all suspects are innocent is a large part of the political problem.

**4.1. A Small Price to Pay?**
Now, even if we accept premise (4), this does not imply that anything untoward is happening. As we said above, it is clear that there are circumstances under which the right to privacy should be breached for a particular person. So, if the machine pattern is enough for a serious suspicion as defined in the relevant law, then a court order should be obtainable. Clearly, one must be aware that total privacy never existed and that breach of privacy is desirable, even for the totally innocent. A good case for this is made, in some detail, in (Etzioni 1999; Bailey 2004, part III).

One other important aspect is how much we trust the agency that is breaching privacy. If it is run by a totalitarian state or a profit-oriented business, we will assume that its purposes are at least partially problematic – but we cannot assume the same for a democratically controlled state structure. This is on of the issues that make the discussion so difficult: in order to evaluate privacy breaches one may think one has to evaluate the motives of the agencies concerned; and little agreement will be reached on this point. It is clear that malicious agencies can do nasty things, and it is also clear that agencies will not admit to any malicious intentions, so I suggest that we try to prove our point even under the assumption of a totally benevolent agency. Note that doing this will imply the demand for significantly reduced rights to profit-oriented businesses who carry out data-mining.

**5. Conclusion: What Should Not Be Done Innocently**
What we have established here is that the use of data-mining techniques constitutes a breach of privacy as soon as a person is involved. That breach is justified, if and only if that person is legally authorized to breach this privacy in this case.

Given that data-mining for security purposes without the involvement of persons is currently impossible, its use always constitutes a breach of privacy. Given that current data-mining is too weak to establish legally sufficient grounds for a violation of privacy, it is ethically wrong.

The more sophisticated defenders of data mining suggest that its problem lies in *how* it is used: "One of the principal reasons for public concern about these tools is that there appears to be no consistent policy guiding decisions when and how to use them." (DeRosa 2004, vii). I have tried to show that *any* use violates a right to privacy – unless it is authorized by the person concerned. Subject-based analysis and matching already presuppose personally identified data, while pattern analysis requires human intervention. All use of blanket data analysis should therefore be subject to the conventional legal controls – which excludes using them on whole populations. We should try to make sure that at least our own state and private agencies do not cross these lines.

**Acknowledgements**

The writing of this paper was carried out mainly during a Stanley J. Seeger Fellowship in Research at Princeton University. I am very grateful for this excellent opportunity. A first version of the paper, entitled "If You Had Nothing to Hide, Would You Still Mind Being Watched by Machines?" was presented at the workshop "Privacy: intercultural perspectives" at ZiF, Bielefeld University, in February 2006. I thank Karsten Weber for the invitation and all participants for the very stimulating discussions at that pleasant meeting. I also thank Gordana Dodig-Crnkovic for very useful written comments.